\documentclass[twoside,letterpaper,twocolumn]{article}

\usepackage{SimBGf}
\usepackage{graphicx,xcolor,subfig}

\title{Quantum computational intelligence for traveltime seismic inversion}

\author{ A. S. Albino (SENAI CIMATEC), O. M. Pires (SENAI CIMATEC), R. F. De Souza (SENAI CIMATEC), P. Nogueira (SENAI CIMATEC), E. G. S. Nascimento (University of Surrey)}

\headauthor{A. S. Albino, O. M. Pires, R. F. De Souza, P. N. Santos, E. G. S. Nascimento}
\headtitle{Quantum computational intelligence for traveltime seismic inversion}

\begin{document}

\maketitle

\begin{abstract}
Quantum computing is in its early stage of implementation. Its capacity has been growing in the last years but its application in several fields of sciences is still restricted to oversimplified problems. In this stage, it is important to identify the situations where quantum computing presents the most promising results to be prepared when the technology is ready to be deployed. The geophysics field has several areas which are limited by the current computation capability, among them the so-called seismic
inversion is one of the most important ones, which are strong candidates to benefit from quantum computing.
In this work, we implement an approach for traveltime seismic inversion through a near-term quantum algorithm based on gradient-free quantum circuit learning. We demonstrate that a quantum computer with thousands of qubits, even if noisy, can solve geophysical problems. In addition, we compared the convergence of the method with the variational quantum algorithms.
\end{abstract}

%%%%%%%%%%%%%%%%%%%%%%%%%%%%%%%%%%%%%%%%%%%%%%%%
%\section{Introdu\c c\~ao \normalfont{(Arial~Bold, 9)}} % ignorar
\section{Introduction}
Retrieving the subsurface parameters through seismic inversion is essential for understanding earthquake dynamics, dam monitoring, mapping ore deposits, and oil reservoirs exploration. However, the application of seismic inversion in all these contexts has required an advance in computational geoscience, especially when involving 3D seismic wave propagation in complex earth environments. For these reasons, any new computational techniques with possibilities to speed up calculations must be explored by the community. Since commercial quantum computing has becoming reality, the geophysical community already started to prospect its possible uses,  for instance . \cite{moradi2018quantum} provide a perspective of the most suitable method to solve the wave equation, highlighting that wave equation solution can be performed through linear system equations so that such systems are appropriate for quantum algorithms being able to solve with exponential speedup.
\cite{greer2020approach} introduced an iterative algorithm to solve PDE-constrained optimization problems on a quantum annealer. They then used this method to invert the velocity parameters of the subsurface using a hybrid approach of classical and quantum computing considering some simple models in their seismic experiments.
\cite{sarkar2018snell} addressed how to explore the potential power of quantum computing in an unusual tomographic challenge, implementing a high-resolution algorithm adequate for quantum computing.
\cite{zhao2016prestack} introduced a new stochastic inversion method based on the Quantum Metropolis-Hastings method to deal with pre-stack seismic inversion problems.
\cite{zhu2020solution} propose a finite difference (FD) method solution based on enhanced quantum particle swarm optimization (QPSO). Their numerical dispersion analysis shows that the optimized FD scheme based on the improved QPSO algorithm has a wider spectral coverage and the accuracy error is controlled within a valid range, which means that the improved QPSO algorithm has better capability to find precise global solutions.

Quantum computing was first proposed by \cite{feynman} as an alternative to solve problems, intractable to classical computing, by performing computational tasks using quantum phenomena. Algorithms that exploit these properties have been developed in the last decades, as well as the construction of quantum devices to perform such tasks, as summarized in \cite{preskill}. Current quantum devices still have a considerable noise rate. These devices are called Noisy Intermediate Scale Quantum (NISQ) computers and handle a relatively small amount of successive operations on an initial state before the state decoherence processes take place. However, given these limitations, with the effort of the scientific community, new noise mitigation techniques and algorithms capable of dealing with NISQ computers called Variational Quantum Algorithms (VQAs) have emerged, as can be seen in \cite{vqa, qaoa, vqe, warmqaoa}. In the works developed by \cite{vqc, vqcadvantage}, it was proved mathematically, and demonstrated through experiments, that NISQ computers can provide an exponential advantage in supervised learning problems.
In \cite{chinesesadvantage}, an experiment was performed to demonstrate a quantum advantage in a mathematically well-defined problem, using a photonic quantum device. More recently, the experiment carried out by \cite{quantumlearning}, demonstrated an exponential advantage in a practical problem of learning properties of quantum systems.

Many of the problems in geophysics can be modeled and solved through solutions of systems of linear equations. Therefore, investigating quantum algorithms capable of performing such tasks is an important step towards the application of quantum computing in geophysics. The work proposed by \cite{hhl}, addresses a way to use quantum computers to solve linear systems with exponential advantage over the best classical method. The method is known as
Harrow Hassidim Lloyd (HHL) Algorithm and takes advantage of Quantum Phase Estimation (QPE) to achieve quantum advantage, which requires error-corrected quantum computers. A different approach to solving linear systems can be seen in \cite{vqls}, where variational circuits are used, in conjunction with the Hadamard Test, to approximate systems solutions. This algorithm is known as the Variational Quantum Linear Solver (VQLS) and, like most variational algorithms, has potential for near-term applications. In the solution of a linear system, $A\vec{x} = \vec{b}$, both algorithms mentioned above have as requirements the decomposition of the matrix $A$ as a linear combination of the set of Pauli Operators and the efficient preparation of the quantum state, which is the normalization of the vector $\vec{b}$. As these algorithms perform amplitude embedding to represent vector b as a quantum state, the number $n$ of qubits needed to implement them scales logarithmically, $n = O(log(B))$, where $ B$ is the number of elements of $\vec{b}$. Such requirements are challenging problems in the current state of quantum computing. In \cite{qea}, an evolutionary strategy was adopted in NISQ devices in order to solve optimization problems. The results showed that the convergence of the evolutionary algorithm is faster compared to the VQAs. In addition, tests were performed on real quantum devices to demonstrate that the algorithm is a strong candidate for near-term applications because it generates low-depth quantum circuits.

In this work, we aim to solve the problem of seismic traveltime inversion using quantum computational intelligence. For this, we model the seismic inversion problem as an unconstrained discrete optimization problem and solve it using an evolutionary algorithm based on gradient-free quantum circuit learning, proposed by \cite{qea}, suitable for use in near-term computers and which has been shown to mitigate the barren plateau phenomena. From the results and the mathematical formulation of the problem, it is possible to demonstrate that NISQ devices with thousands of qubits can be used to solve systems of linear equations. Also, the method could be used to solve differential equations, since they can be discretized using the finite difference method. Although the limitations of the quantum computing applications in real geosciences problems are still enormous, the idea here is to show that it is possible to advance in seismic inversion using quantum computing, even applying these techniques in easy experiments.

%\section{Theory}

\section{Methodology}

\subsection{Theoretical Foundations}

%\subsubsection{Quantum Circuit Learning}

Evolutionary computing is a set of algorithms inspired by biological behaviors which optimize a given problem by iteratively working with a population of solutions and selecting the best ones over a number of generations.
The work developed by \cite{qea} shows an evolutionary quantum optimization strategy analogous to the Neuroevolution of Augmenting Topologies (NEAT). In this quantum evolutionary scheme, candidates are quantum circuits and the cost function is the expected value of the cost hamiltonian evaluated from the measurements in the quantum circuits.

The objective function can be written with respect to some target Hamiltonian $H$ as 
\begin{equation}\label{exp-value}
    f(U) = \langle \psi_0 |U^\dagger H U|\psi_0 \rangle  ,
\end{equation}
whose variable $U$ is a quantum circuit that evolves at each generation. The optimal quantum circuit is given by $U_{opt} = U_0U_1...U_n$ if the optimization is carried out over a number of $n$ generations. The expected value of $H$ can also be written in numerical form as
\begin{equation}
    \langle \psi_0 |U^\dagger H U|\psi_0 \rangle = \sum_{x=0}^{}p_xE_x
\end{equation}
where $p_x$ is the measurement probability of an eigenstate of the computational base $|x\rangle$ being measured and $E_x$ its associated cost. Therefore, the computational cost of the algorithm will be related to the amount of measurements made on the state $|\psi\rangle$, being generally $O(poly)$. Since eigenstates are represented in the computational basis, the cost hamiltonian $H$ should be written as a Quadratic Unconstrained Binary Optimization (QUBO) or as an Ising Model.

During each generation, an amount of $\lambda$ circuits are created from a parent circuit through a process of mutation. The offspring candidates have their loss functions evaluated and then undergo a selection process, in which the best individual becomes the parent of the next generation. By the end of the algorithm, the best individual of the last generation is returned as the final answer.

The mutation strategy is described by four operations that are performed randomly to the circuit. The operations are:
\begin{itemize}
    \item \textbf{insert} - insertion of single or two-qubit rotation gates. The set of single qubit gates which can be applied is $\{Rx(\theta), Ry(\theta), Rz(\theta)\}$, while the set of two-qubit gates is $\{cRx(\theta), cRy(\theta), cRz(\theta)\}$. The rotation angles, $\theta$, are hyperparameters and should be initialized randomly;
    \item \textbf{delete} - delete one of the already applied gates. The effect of this operation, in addition to improving the search space and finding better solutions, is also to reduce the depth of the quantum circuit, an important task in the case of noisy quantum computers.;
    \item \textbf{modify} - modify the rotation angle, $\theta$, of an already applied single-qubit or two-qubit gate. This modification is done with random values of $\theta$;
    \item \textbf{swap} - Flip the control and target of a two-qubit gate in a given position. It has the same effect of a deletion followed by an insertion.
\end{itemize}
Each operation has a probability of happening tied to the problem being optimized. This procedure is suitable for near-term applications, since the number of operations of the final circuit is relatively small. In addition, as it is a gradient-free optimization method that does not use ansatz, it has been shown to mitigate the undesired effect of barren plateau.

%%%%%%%%%%%%%%%%%%%%%%%%%%%%%%%%%%%%%%%%%%%%%%%%%%%%%%%%
%\section{Metodologia/Problema Investigado \normalfont{(Arial~Bold, 9)}} % ignorar
\subsection{Seismic inversion as a discrete optimization problem}

\begin{figure}[h!]
\centering
\includegraphics[width=\linewidth]{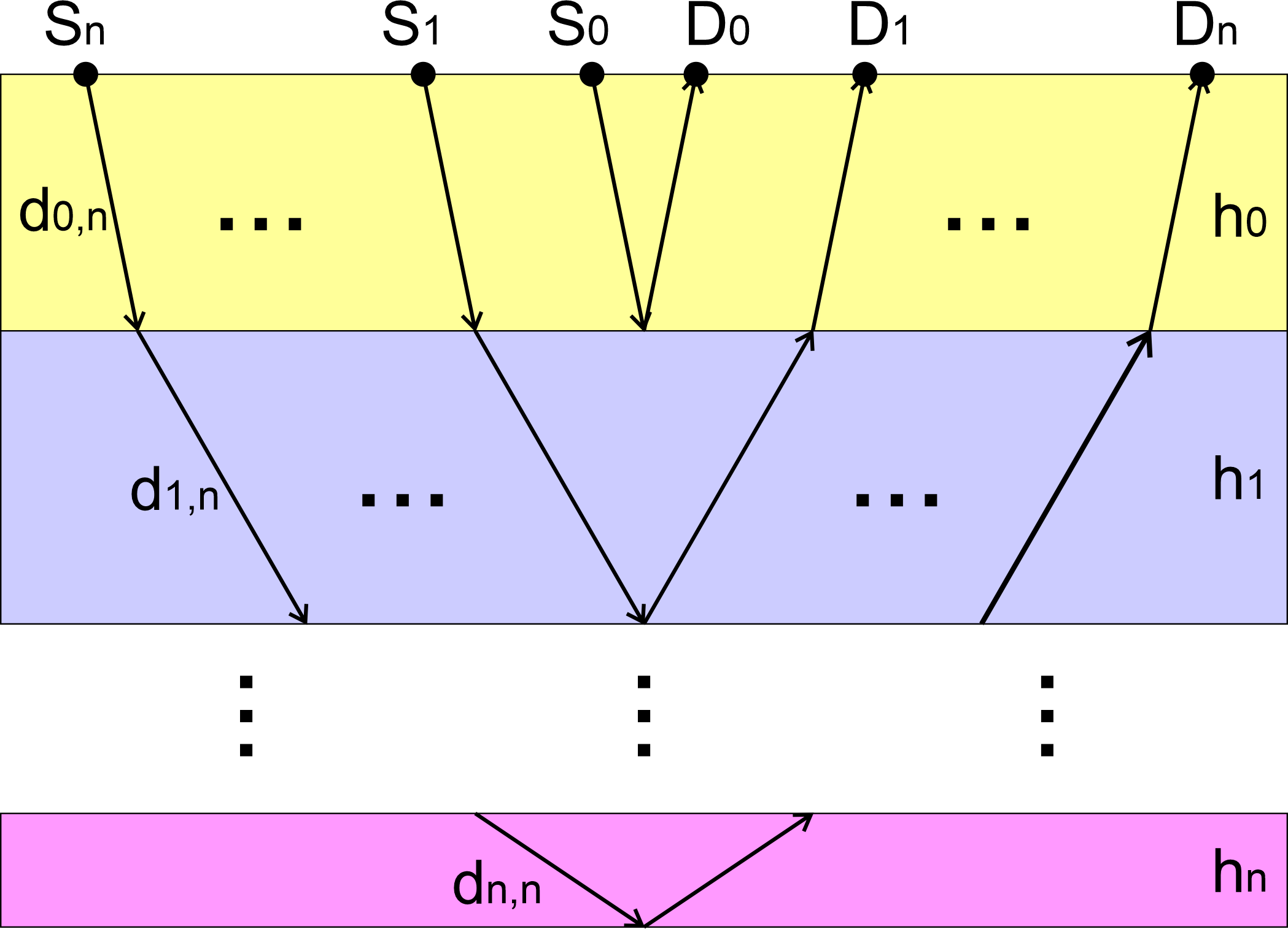}
\caption{Subsurface model with $n$ parallel layers. The layers are separated by a distance $h_i$. The direction of propagation of the wave is from the sources $S_i$ to the detectors $D_i$, passing through the subsurface with different propagation distances $d_{ij}$ in each of them. The different colors represent wave velocities in each subsurface layer, which must be determined.}
\label{fig:simodel}
\end{figure}

The problem of determining the subsurface wave propagation velocities $\vec{v}$ from observed travel time data, $\vec{t}$, can be written using the system of linear equations
\begin{equation}
    D\vec{s} = \vec{t}
\end{equation}
whose matrix $D$ is the upper diagonal and represents the distances, $d_{ij}$ (see Fig. \ref{fig:simodel}), of wave propagation in each subsurface layer and $\vec{s}$ is the slowness vector, whose elements are given by $s_i = 1/v_i$.

In order to use the Evolutionary Circuit Learning method described above, the linear systems problem must be written as the unconstrained optimization problem.

\begin{equation}\label{lspot}
    f(\vec{s}) = \|D\vec{s} - \vec{t}\|^2.
\end{equation}

The value of $f(\vec{s})$ is the square of the norm of the difference between $D\vec{s}$ and $\vec{t}$, and the null result in case of an exact solution. However, since $s_i$ is a real number, and measurements performed on a quantum state, $|\psi\rangle$, give us results on the computational basis, $\{0.1\}$, we must write Eq. \ref{lspot} as an objective function of binary variables, $x_i$, through the approximation 
\begin{equation}
    s_i = \sum_{r=-R}^{R}x_i 2^r.
\end{equation}
In this way, we can write the resulting objective function, $H = f(\vec{x})$. As $H$ is a QUBO, it can also be solved using Variational Quantum Algorithms (VQAs), as in the seismic inversion model proposed by \cite{qcseismiccimatec}. In these approaches, the mandatory number of qubits for the solution of a linear system of order $M \times M$ is $n = O(RM)$, where $R$ is the number of bits of the approximation.

%%%%%%%%%%%%%%%%%%%%%%%%%%%%%%%%%%%%%%%%%%%%%%%%%%%%%%%%
%\section{Resultados \normalfont{(Arial~Bold, 9)}} % ignorar
\section{Results and Discussion}

We simulated our circuits using the Qiskit statevector simulation in a Intel Core i7-8550U CPU. We calculated the seismic inversion for 3, 4 and 6 layers. For each number of layers we repeated the evolutionary circuit learning 10 times and we returned the mean of the best result of all runs as the final result. Each run had 300 generations, the number of parents $\mu = 1$ and the number of offspring $\lambda = 4$.

In all instances, synthetic slowness models were created with 3-bit approximation to integer numbers. Since the number of qubits scales as $n=O(RM)$, for 3, 4 and 6 speed models, the simulations were performed using 9, 12 and 18 qubits, respectively. 

\begin{figure}

         \centering
         \includegraphics[width=\linewidth]{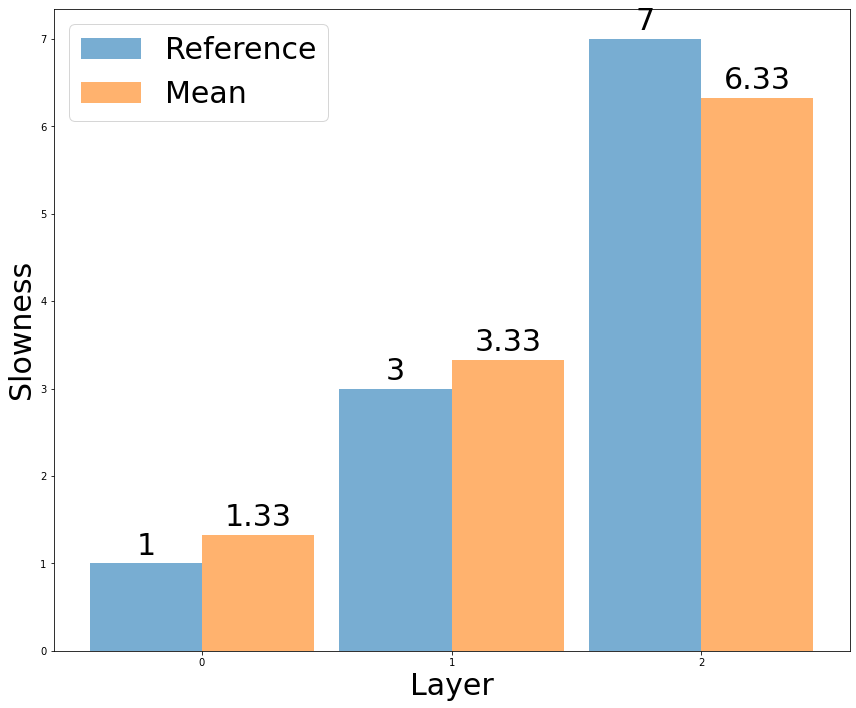}
         %\caption{Three layers.}
         
         %\label{fig:y equals x}

     %\hfill

         \centering
         \includegraphics[width=\linewidth]{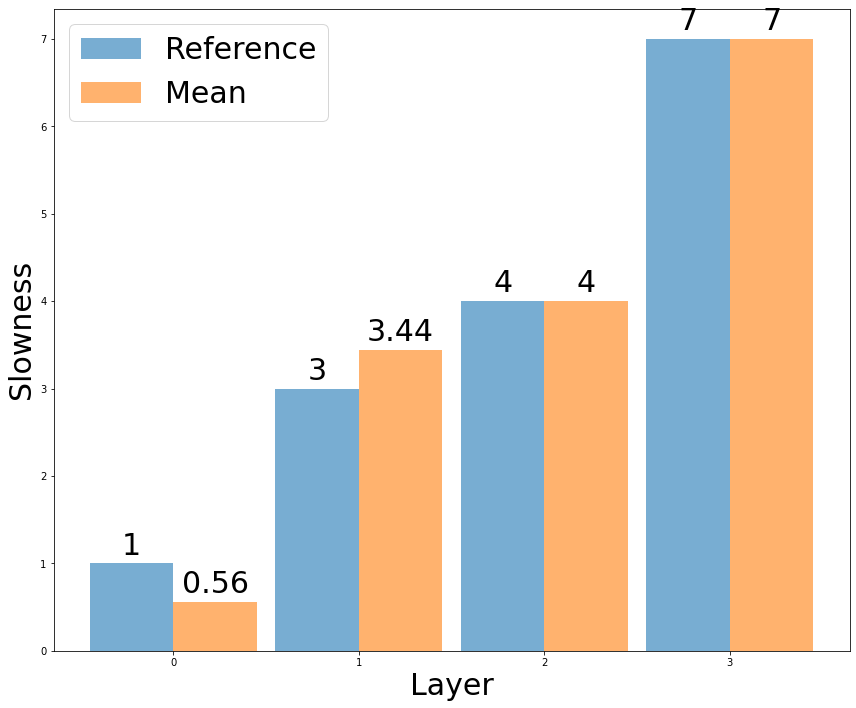}
         %\caption{Four layers.}
         %\label{fig:three sin x}

     %\hfill

         \centering
         \includegraphics[width=\linewidth]{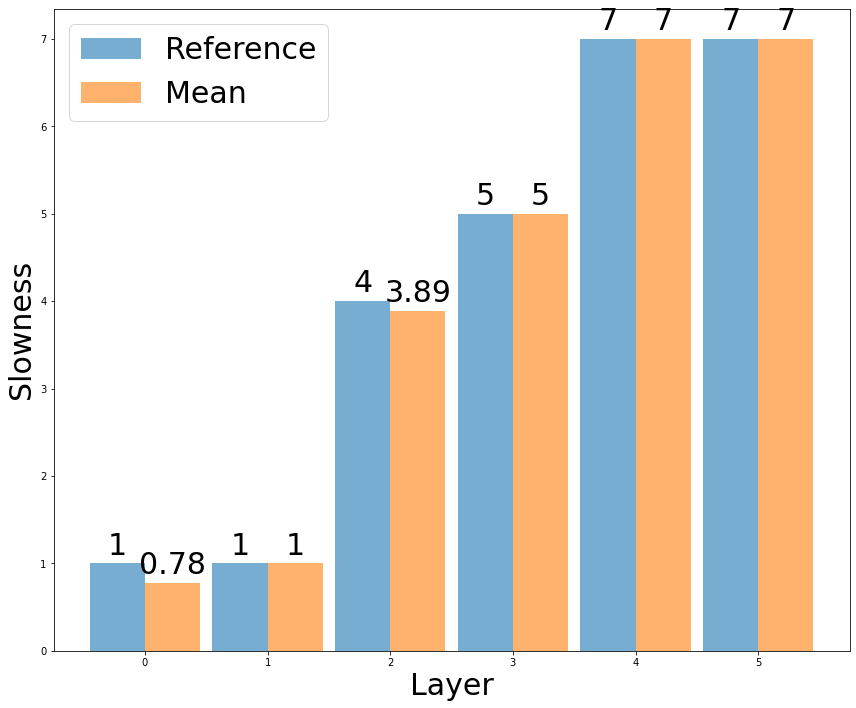}
         %\caption{Six layers.}
         %\label{fig:five over x}

        \caption{Comparison histogram between the reference values (synthetic 3-bit slowness model) and the average values of the results obtained in 10 rounds of the algorithm. The figure shows three histograms, the results being for the models with 3, 4 and 6 layers, in this order (from top to bottom).}
        \label{fig:three graphs}
\end{figure}

To compare the synthetic velocity model and the one calculated by the quantum algorithm, a statistical treatment was performed so that, for each instance, the algorithm was run 10 times and the probability distributions for different generations were stored. An average was obtained over the results of the 10 runs, which can be seen in Fig. \ref{fig:three graphs}

\begin{figure}

         \centering
         \includegraphics[width=\linewidth]{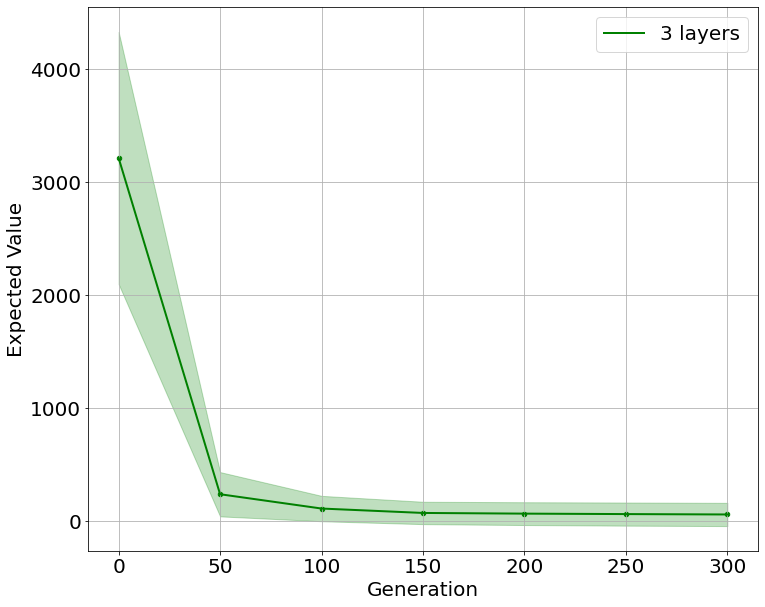}
         %\caption{Three layers.}
         
         %\label{fig:y equals x}

     %\hfill

         \centering
         \includegraphics[width=\linewidth]{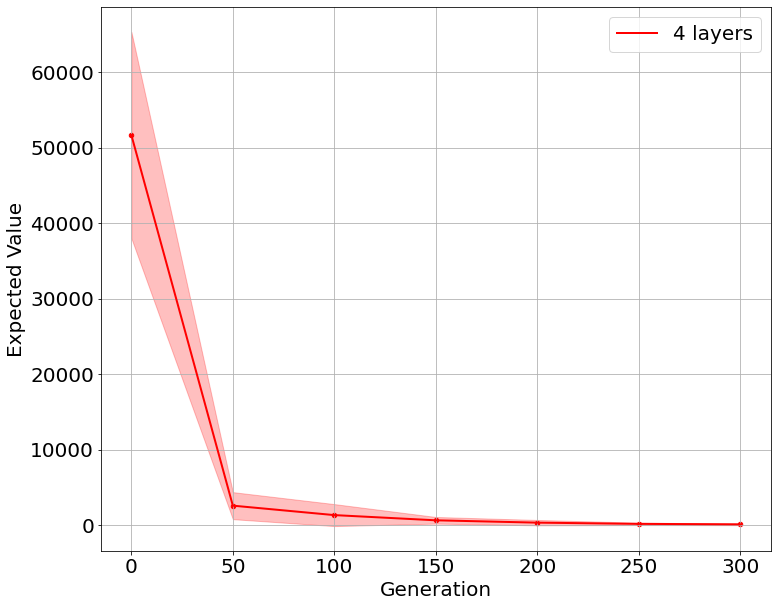}
         %\caption{Four layers.}
         %\label{fig:three sin x}

     %\hfill

         \centering
         \includegraphics[width=\linewidth]{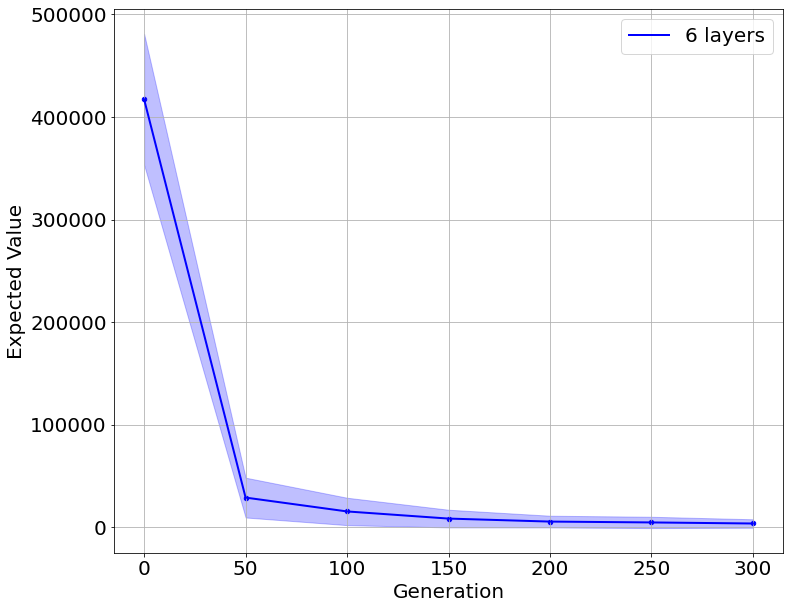}
         %\caption{Six layers.}
         %\label{fig:five over x}

        \caption{Convergence graphs for observing the cost (expected value) as a function of the number of generations (we analyzed for generations 0, 50, 100, 150, 200, 250 and 300). We observed that all instances followed a similar pattern of approximation to the optimal value (zero). The shaded values represent the standard deviation and the lines are average values of the cost function for the 10 runs of the algorithm.}
        \label{fig:3graphsexpvalue}
\end{figure}

The convergence of the method (See Fig. \ref{fig:3graphsexpvalue}) was faster when compared to the variational algorithms, which in general use static ansatz and gradient-based optimizers. The convergence metric was the objective function itself, the expected value of $H$, which gives a better idea of the algorithm scalability. As the number $n$ of qubits increases, the number of possible solutions in superposition tends to grow exponentially, but as the number of measurements performed was polynomial, the number of solutions measured is polynomial or quasilinear. Therefore, for larger instances, the expected value tends to be consequently much larger, even with good solutions in the final distribution.

\begin{figure}[h!]
\centering
\includegraphics[width=\linewidth]{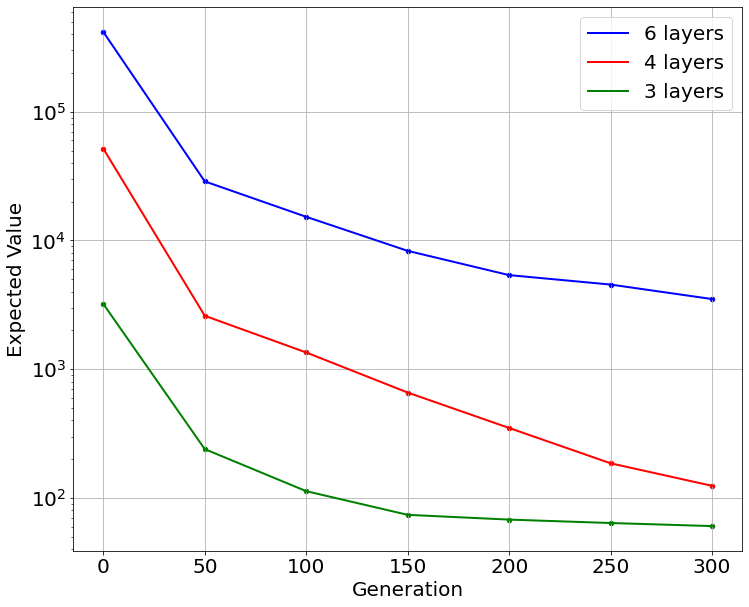}
\caption{Logarithmic scale of algorithm convergence comparing the three instances implemented in the simulations. The difference in cost for different layers is due to the increase in the number of measured eigenstates, consequently increasing the expected value of $H$.}
\label{fig:convlog}
\end{figure}

%%%%%%%%%%%%%%%%%%%%%%%%%%%%%%%%
%\section{Discuss\~ao e Conclus\~oes \normalfont{(Arial~Bold, 9)}} % ignorar
\section{Conclusions}
Quantum computing offers potential computational advantage due to the possibility of creating and manipulating states in a higher-order space, the Hilbert space. Demonstrations of quantum advantage have been demonstrated in recent years, and advances in this area have grown rapidly in light of the impact that technology consolidation can have. Due to the computational cost of the geosciences problems addressed and the impact if there are more efficient ways to solve them, quantum computing emerges as a strong candidate to accelerate these processes, even with noisy quantum computers. 

In this work, we implemented a quantum evolutionary algorithm to tackle the traveltime seismic inversion problem modeled as a QUBO for three, four and six parallel layers. Our simulation experiments showed great rate of convergence and the method compared to variational quantum algorithms, finding results close to the exact solution.  The investigated algorithm demonstrated that it generates low-depth quantum circuits even in the final generations, which makes it a good candidate to solve near-term optimization problems.

The main findings and contributions of this work were the use of a quantum computational intelligence method for gradient-free and ansatz-free optimization where the quantum circuit depth grows iteratively with the evolution of generations. Therefore, experiments have shown that systems of linear equations and differential equations with relevant applications can be solved with this method with quantum devices with a few thousand qubits, which has already been promised for the next few years.

%%%%%%%%%%%%%%%%%%%%%%%%%%%%%%%%
%\section{Agradecimentos \normalfont{(Arial~Bold, 9)}} % ignorar
\section{Acknowledgements}
Acknowledgements to the Supercomputing Center for Industrial Innovation (CS2i), the Reference Center for Artificial Intelligence (CRIA), and the Latin American Quantum Computing Center(LAQCC), all from SENAI CIMATEC. R.F de Souza  would like to thank the financial support provided by Financiadora de Estudos e Projetos (FINEP) - grant number 0113032700 and Ministério da Ciência, Tecnologia e Inovações (MCTI).

%\newpage % omitir se necessario

%%%%%%%%%%%%%%%%%%%%%%%%%%%%%%%%
% Referencias 

%\renewcommand{\refname}{Refer\^encias \normalfont{(Arial~Bold, 9)}} % ignorar
%\nocite{*} % ignorar

\bibliographystyle{SimBGf}
\bibliography{refs}

\begin{thebibliography}{19}
\providecommand{\natexlab}[1]{#1}
\providecommand{\url}[1]{\texttt{#1}}
\providecommand{\urlprefix}{URL }
\expandafter\ifx\csname urlstyle\endcsname\relax
  \providecommand{\doi}[1]{doi:\discretionary{}{}{}#1}\else
  \providecommand{\doi}{doi:\discretionary{}{}{}\begingroup
  \urlstyle{rm}\Url}\fi

\bibitem[{Albino et~al.(2022)}]{qcseismiccimatec}
Albino, A. et~al., 2022. Employing gate-based quantum computing for travel time
  seismic inversion, vol. 2022(1): 1--5,
  \doi{https://doi.org/10.3997/2214-4609.2022.80007}.

\bibitem[{Bravo-Prieto et~al.(2019)}]{vqls}
Bravo-Prieto, C. et~al., 2019. Variational quantum linear solver,
  \doi{10.48550/ARXIV.1909.05820},
  \urlprefix\url{https://arxiv.org/abs/1909.05820}.

\bibitem[{Cerezo et~al.(2021)}]{vqa}
Cerezo, M. et~al., 2021. Variational quantum algorithms, Nature Reviews
  Physics, vol.~3(9): 625--644, \doi{10.1038/s42254-021-00348-9}.

\bibitem[{Deshpande et~al.(2022)}]{chinesesadvantage}
Deshpande, A. et~al., 2022. Quantum computational advantage via
  high-dimensional gaussian boson sampling, Science Advances, vol.~8(1),
  \doi{10.1126/sciadv.abi7894}.

\bibitem[{Egger et~al.(2021)Egger, Mare{\v{c} }ek \& Woerner}]{warmqaoa}
Egger, D.~J., Mare{\v{c} }ek, J. \& Woerner, S., 2021. Warm-starting quantum
  optimization, Quantum, vol.~5: 479, \doi{10.22331/q-2021-06-17-479}.

\bibitem[{Feynman(1982)}]{feynman}
Feynman, R., 1982. Simulating physics with computers, International Journal of
  Theoretical Physics, vol.~21(1), \doi{10.1007/BF02650179}.

\bibitem[{Franken et~al.(2020)}]{qea}
Franken, L. et~al., 2020. Quantum circuit evolution on nisq devices,
  \doi{10.48550/ARXIV.2012.13453},
  \urlprefix\url{https://arxiv.org/abs/2012.13453}.

\bibitem[{Greer \& O’Malley(2020)}]{greer2020approach}
Greer, S. \& O’Malley, D., 2020. An approach to seismic inversion with
  quantum annealing, in: SEG Technical Program Expanded Abstracts 2020, Society
  of Exploration Geophysicists, 2845--2849.

\bibitem[{Harrow et~al.(2009)Harrow, Hassidim \& Lloyd}]{hhl}
Harrow, A.~W., Hassidim, A. \& Lloyd, S., 2009. Quantum algorithm for linear
  systems of equations, Physical Review Letters, vol. 103(15),
  \doi{10.1103/physrevlett.103.150502}.

\bibitem[{Havl{\'{\i}}{\v{c}}ek et~al.(2019)}]{vqc}
Havl{\'{\i}}{\v{c}}ek, V. et~al., 2019. Supervised learning with
  quantum-enhanced feature spaces, Nature, vol. 567(7747): 209--212,
  \doi{10.1038/s41586-019-0980-2}.

\bibitem[{Huang et~al.(2022)}]{quantumlearning}
Huang, H.-Y. et~al., 2022. Quantum advantage in learning from experiments,
  Science, vol. 376(6598): 1182--1186, \doi{10.1126/science.abn7293}.

\bibitem[{Liu et~al.(2021)Liu, Arunachalam \& Temme}]{vqcadvantage}
Liu, Y., Arunachalam, S. \& Temme, K., 2021. A rigorous and robust quantum
  speed-up in supervised machine learning, Nature Physics, vol.~17(9):
  1013--1017, \doi{10.1038/s41567-021-01287-z}.

\bibitem[{Moradi et~al.(2018)Moradi, Trad \& Innanen}]{moradi2018quantum}
Moradi, S., Trad, D. \& Innanen, K.~A., 2018. Quantum computing in geophysics:
  Algorithms, computational costs, and future applications, in: 2018 SEG
  International Exposition and Annual Meeting, OnePetro.

\bibitem[{Peruzzo et~al.(2014)}]{vqe}
Peruzzo, A. et~al., 2014. A variational eigenvalue solver on a photonic quantum
  processor, Nature Communications, vol.~5(1), \doi{10.1038/ncomms5213}.

\bibitem[{Preskill(2021)}]{preskill}
Preskill, J., 2021. Quantum computing 40 years later,
  \doi{10.48550/ARXIV.2106.10522},
  \urlprefix\url{https://arxiv.org/abs/2106.10522}.

\bibitem[{Sarkar \& Levin(2018)}]{sarkar2018snell}
Sarkar, R. \& Levin, S.~A., 2018. Snell tomography for net-to-gross estimation
  using quantum annealing, in: 2018 SEG International Exposition and Annual
  Meeting, OnePetro.

\bibitem[{Zhao \& Zhang(2016)}]{zhao2016prestack}
Zhao, C. \& Zhang, G., 2016. The prestack seismic stochastic inversion based on
  quantum metropolis--hastings method, in: SEG Technical Program Expanded
  Abstracts 2016, Society of Exploration Geophysicists, 562--566.

\bibitem[{Zhou et~al.(2020)Zhou, Wang, Choi, Pichler \& Lukin}]{qaoa}
Zhou, L., Wang, S.-T., Choi, S., Pichler, H. \& Lukin, M.~D., 2020. Quantum
  approximate optimization algorithm: Performance, mechanism, and
  implementation on near-term devices, Phys. Rev. X, vol.~10: 021067,
  \doi{10.1103/PhysRevX.10.021067}.

\bibitem[{Zhu et~al.(2020)Zhu, Wang, Yu \& Li}]{zhu2020solution}
Zhu, M., Wang, Z., Yu, D. \& Li, Y., 2020. Solution of finite difference method
  based on improved quantum particle swarm optimization, in: SEG International
  Exposition and Annual Meeting, OnePetro.

\end{thebibliography}

%%%%%%%%%%%%%%%%%%%%%%%%%%%%%%%%
% Figuras grandes

\newpage
\onecolumn
%
%\begin{figure}[ht]
%\centering
%\includegraphics[width=\linewidth]{figura4.png}
%\caption{Qqwertqwertqwertqwertqwertqwertqwertqwertqwertqwertqwertqwertqwertqwertqwert.}
%\label{fig:exemplo2}
%\end{figure}

\end{document}